\documentclass[a4paper, 12pt]{article}
\begin{document}
\title{Remark on the core/halo model of Bose-Einstein correlations
in multiple particle production processes
\thanks{Supported in part by the KBN grant 2P03B 093 22}
}
\author{K.Zalewski\\
M.Smoluchowski Institute of Physics,
 Jagellonian University,\\ Cracow, Reymonta 4, 30 059 Poland,\\ e-mail:
zalewski@th.if.uj.edu.pl \\ and \\ Institute of Nuclear Physics
PAN, Cracow
 }
\maketitle
\begin{abstract}
The core/halo model describes the Bose-Einstein correlations in
multihadron production taking into account the effects of
long-lived resonances. The model contains the combinatorial
coefficients $\alpha_j$ which were originally calculated from a
recurrence relation. We show that $\alpha_j$ is the integer
closest to the number $j!/e$.
\end{abstract}
Bose-Einstein correlations.\vspace{1cm}

Bose-Einstein correlations in multiple particle production
processes are much studied in order to get information about the
interaction regions and about the hadronization processes. For
detailed reviews see e.g. \cite{WIH}, \cite{WEI}, \cite{CSO1}. The
starting point is, usually, the factorizeable approximation (see
e.g. \cite{KAR}, \cite{BIK}), where the $n$-particle density
matrix for $n$ undistinguishable particles is:

\begin{equation}\label{}
  \rho(\textbf{p}_1,\ldots,\textbf{p}_n;\textbf{p}_1',\ldots,\textbf{p}_n') =
  \frac{1}{n!}\sum_{P,Q}\prod_{j=1}^n \rho(\textbf{p}_{j_P};\textbf{p}_{j_Q}'),
\end{equation}
$\rho(\textbf{p};\textbf{p}')$ is some single particle density
matrix and the summation is over all the permutations of the
indices of $\textbf{p}$ and $\textbf{p}'$. The corresponding
unsymmetrized density matrix is:

\begin{equation}\label{}
\rho^U(\textbf{p}_1,\ldots,\textbf{p}_n;\textbf{p}_1',\ldots,\textbf{p}_n')
= \prod_{j=1}^n \rho(\textbf{p}_j;\textbf{p}_j').
\end{equation}
The quantities usually presented are the $n$-body correlation
functions

\begin{equation}\label{}
  C_n(\textbf{p}_1,\ldots,\textbf{p}_n) =
  \frac{\rho(\textbf{p}_1,\ldots,\textbf{p}_n;\textbf{p}_1,\ldots,\textbf{p}_n)}
  {\rho^U(\textbf{p}_1,\ldots,\textbf{p}_n;\textbf{p}_1,\ldots,\textbf{p}_n)}.
\end{equation}
As seen from the definitions $C_n(\textbf{p},\ldots,\textbf{p}) =
n!$. There is a number of difficulties to check this prediction.
On the experimental side, a pair of momenta, say $\textbf{p}_i$
and $\textbf{p}_j$, can be reliably measured only if  $Q_{ij} =
\sqrt{-(p_i - p_j)^2}$ exceeds some $Q_{min}$. At present it is
difficult to go with $Q_{min}$ below some $(5 - 10)$ MeV (see e.g.
\cite{CHZ} and references given there). This is important, because
the very small $Q$ region is expected to contain narrow peaks due
to long-lived resonances \cite{GRA}. There are other difficulties
in the small $Q$ region: Coulomb corrections, misidentified
particles, perhaps effects of coherence. A model which assumes
that these other factors can be either corrected for or neglected
is the core/halo model\footnote{Let us note, however an attempt to
include partial coherence into this model \cite{CSO2}} \cite{CLZ},
\cite{CSO3}. According to this model for a group of identical
particles close to each other in momentum space, but not so close
as not to be resolved experimentally, the single particle density
matrix elements $\rho(\textbf{p}_{j_P};\textbf{p}'_{j_Q})$ reach
for $j_Q \neq j_P$ a common limit
$f(\textbf{p})\rho(\textbf{p};\textbf{p})$, where p is some
average momentum of the particles in the group. Of course for $j_Q
= j_P$ the matrix element in the numerator cancels with the
corresponding matrix element in the denominator. Thus the
correlation function extrapolated from the region accessible
experimentally to the point $\textbf{p}_1=\ldots,=\textbf{p}_n$ is

\begin{equation}\label{}
  C_n^{extr}(\textbf{p},\ldots,\textbf{p}) = \sum_{j=0}^n \left(\begin{array}{c}n \\j \ \end{array}\right) \alpha_j f(\textbf{p})^j
\end{equation}
where $\alpha_j$ is the number of permutations of $j$ elements
where no element keeps its place.

Let us note the identity

\begin{equation}\label{recurr}
  k! = \sum_{j=0}^k \left(\begin{array}{c}
    k \\
    j \
  \end{array}\right)\alpha_j,
\end{equation}
following from the remark that every permutation of $k$ elements
can be characterized by the number $j$ of elements which changed
their places and that the number of choices of these elements is
$\left(\begin{array}{c}k \\j \ \end{array}\right)$. The formula
can be used to calculate the coefficient $\alpha_j$ when all the
coefficients $\alpha_k$ with indices $k < j$ are known
\cite{CSO3}. In this note we derive a simpler formula for the
coefficients $\alpha_n$. Several derivations of this result can be
found in mathematical textbooks. Here we use the idea of the proof
from ref. \cite{GKP}.

Let us multiply both sides of (\ref{recurr}) by
$(-1)^k\left(\begin{array}{c}n \\k \
\end{array}\right)$ and sum over $k$ from zero to $n$. On the
left-hand side we get

\begin{equation}\label{}
  n!\sum_{k=0}^n\frac{(-1)^k}{(n-k)!} = n!(-1)^n \sum_{k=0}^n\frac{(-1)^k}{k!}
\end{equation}
and on the right-hand side

\begin{eqnarray}\label{}
\sum_{k=0}^n(-1)^k\left(\begin{array}{c}n \\k \ \end{array}\right)
\sum_{j=0}^n\left(\begin{array}{c}k \\j \
\end{array}\right)\alpha_j = \nonumber \\\sum_{j=0}^n\alpha_j(-1)^j \left(\begin{array}{c}n \\j \
\end{array}\right)\sum_{k=j}^n \left(\begin{array}{c}n-j \\k-j \
\end{array}\right)(-1)^{k-j} = (-1)^n\alpha_n,
\end{eqnarray}
where the second equality follows from the remark that the sum
over $k$ yields $1$ for $j=n$ and $(1-1)^{n-j} = 0$ for $j<n$.
Comparing the two sides one finds

\begin{equation}\label{}
  \alpha_n = n!\sum_{k=0}^n\frac{(-1)^k}{k!}.
\end{equation}
The coefficient of $n!$ tends to $e^{-1}$ when $n$ increases. It
is, however, an alternating series with monotonically decreasing,
non-zero terms. For such series the sum of the first $n$ elements
approximates the limit with an error less than the absolute value
of the first rejected term. Thus
\begin{equation}\label{}
  \left|\alpha_n - \frac{n!}{e}\right| < \frac{1}{n+1}
\end{equation}
and $\alpha_n$, for $n>0$, can be calculated as the integer
closest to $n!/e$.

 The author thanks T. Cs\"org\"o for a stimulating exchange of
 e-mails as well as R. Wit and M. Zalewski for discussions on the
 mathematical aspects of the problem.

\end{document}